\documentclass[a4paper,11pt]{article} 
\usepackage{amssymb,amsmath,amsfonts,makeidx,placeins,pbox,multirow}
\usepackage{graphicx,rotate,subcaption,color,slashed,cite,caption,epstopdf,verbatim,url,multirow}
\usepackage{longtable,tabu}  \usepackage{array}
\newcolumntype{P}[1]{>{\centering\arraybackslash}p{#1}}
\usepackage[colorlinks=true, linkcolor=magenta, urlcolor=blue,
citecolor=blue]{hyperref} \usepackage[utf8x]{inputenc}

\numberwithin{equation}{section}
\long\def\change#1!!!#2!!!{\color{red}#1 \color{blue}#2 \color{black}}
\evensidemargin=\oddsidemargin \topmargin -1.0cm 
\begin{document}
	%\maketitle
	
	\begin{center}
		
		{\Large \bf Probing the Higgs boson through Yukawa force} \\
		\vspace*{0.5cm} {\sf Avik
			Banerjee\,\footnote{avik.banerjeesinp@saha.ac.in}, ~Gautam
			Bhattacharyya\,\footnote{gautam.bhattacharyya@saha.ac.in}} \\
		\vspace{10pt} {\small } {\em Saha Institute of Nuclear Physics, HBNI,
			1/AF Bidhan Nagar, Kolkata 700064, India} \normalsize
	\end{center}
	
	%%%%%%%%%%%%%%%%%%%%%%%%%%%%%%%%%%%%%%%%%%%%%%%%%%%%%%%%%%%%%%%%%%%%%%%%%%%ABSTRACT%%%%%%%%%%%%%%%%%%%%%%%%%
	
	\begin{abstract}
		
		The ATLAS and CMS collaborations of the LHC have observed that the
		Higgs boson decays into the bottom quark-antiquark pair, and have also
		established that the Higgs coupling with the top quark-antiquark pair
		is instrumental in one of the modes for Higgs production. This
		underlines the discovery of the Yukawa force at the LHC. We
		demonstrate the impact of this discovery on the Higgs properties that
		are related to the dynamics of electroweak symmetry breaking. We show
		that these measurements have considerably squeezed the allowed window
		for new physics contributing to the Higgs couplings with the weak
		gauge bosons and the third generation quarks. The expected constraints
		at the HL-LHC and future Higgs factories are also shown. We project these constraints on the
		parameter space of a few motivated scenarios beyond the Standard
		Model. We pick them under two broad categories, namely, the composite
		Higgs and its RS dual, as well as various types of multi-Higgs
		models. The latter category includes models with singlet scalars, Type
		I, II and BGL-type two-Higgs doublet models, and models with scalar
		triplets {\em \`a la} Georgi and Machacek.
		
	\end{abstract}
	
	%%%%%%%%%%%%%%%%%%%%%%%%%%%%%%%%%%%%%%%%%%%%%%%%%%%%%%%%%%%%%%%%%%%%%%%%%%%%%%%%%%%%%%%%%%%%%%%%%%%%%%%%%%%%
	
	\bigskip
	
	%%%%%%%%%%%%%%%%%%%%%%%%%%%%%%%%%%%%%%%%%%%%%%%%%%%%%%%%%%%%%%%%%%%%%%%%%%MAIN DOCUMENT%%%%%%%%%%%%%%%%%%%%%
	
	\section{Introduction}
	\label{intro}
	
	Since the discovery of the Higgs boson in 2012 at the CERN Large
	Hadron Collider (LHC) \cite{Aad:2012tfa,Chatrchyan:2012xdj}, one of
	the most notable achievements by the ATLAS and CMS Collaborations has
	been the measurements of the Yukawa force between the Higgs boson
	($h$) and the third generation quarks ($t$ and $b$). Although Yukawa
	interaction was postulated long back in the context of the
	pion-nucleon scattering, advent of Quantum Chromodynamics showed that
	it is but an artefact of the strong gauge force.  Do the present
	measurements of $hb\bar{b}$ \cite{Sirunyan:2018kst,Aaboud:2018zhk} and
	$ht\bar{t}$ \cite{Sirunyan:2018hoz,Aaboud:2018urx} couplings
	constitute a discovery of a fundamental Yukawa force, or, it is again
	a low energy manifestation of some unknown UV dynamics? Even if the
	Higgs boson is an elementary object, is it the only neutral scalar
	that Nature offered us? Precision measurements of these Yukawa
	couplings can shed important light on both these questions. In the
	Standard Model (SM), the Yukawa couplings are precisely known in terms
	of the fermion masses. Any departure would indicate physics beyond the
	SM (BSM) triggering electroweak symmetry breaking
	\cite{Bar-Shalom:2018rjs,Arroyo-Urena:2020fkt,Dicus:2004rg,Dicus:2005ku}.  In this paper, we
	review the status of some BSM physics in the light of the LHC data
	armed with the new measurements of the Yukawa forces. Since the flavor
	changing couplings of the 125 GeV Higgs boson are already too
	constrained, increasingly precise measurements of the flavor diagonal
	couplings at the LHC are essential to probe the Yukawa structure. In
	order to quantify the BSM window we employ a $\chi^2$-analysis using
	the Higgs signal strength data from the ATLAS and CMS Collaborations.
	The Run 2 data \cite{Aad:2019mbh,CMS:2020gsy} with improved
	measurements of the $hb\bar{b}$ and $ht\bar{t}$ couplings, compared to
	what Run 1 could achieve
	\cite{Khachatryan:2016vau,Azatov:2012bz,Ellis:2013lra,Einhorn:2013tja,Pomarol:2013zra,deBlas:2014ula,Bernon:2014vta},
	penetrate rather deep into the BSM parameter space, leading to new
	constraints.  To show the future prospects, we give projections for
	these measurements at the high luminosity runs of the LHC (HL-LHC)
	\cite{Cepeda:2019klc}. We also comment on the expected sensitivities
	of the Higgs coupling measurements in the proposed International
	Linear Collider (ILC) \cite{Dawson:2013bba,Asner:2013psa}, and
	circular $e^+e^-$ colliders -- CEPC
	\cite{Ge:2016tmm,Gu:2017ckc,An:2018dwb} and FCC-ee
	\cite{Abada:2019lih}. For our purpose, we employ a simple model
	independent phenomenological Lagrangian up to two-derivative order,
	which essentially captures modification of the Higgs couplings
	\cite{Falkowski:2013dza,Buchalla:2015qju,Cepeda:2019klc}.  We also
	translate the limits of our model independent parameter space to the
	space of two broad BSM categories, namely, the composite Higgs and
	multi-Higgs models, under the guise of their different {\em avatars},
	which can address the questions raised above.

	%%%%%%%%%%%%%%%%%%%%%%%%%%%%%%%%%%%%%%%%%%%%%%%%%%%%%%%%%%%%%%%%%%%%%%%%%%%%%%%%%%%%%%%%%%%%%%%%%%%%%%%%%%%% 
	\section{Theoretical framework}
	\label{th_frmwrk}
	
	Broadly speaking, two types of BSM physics can modify the Higgs boson
	couplings:
	\begin{itemize}
		\item Mixing with other spin-0 bosons can alter the Higgs
		couplings. Examples of this type are found in models with additional
		$\rm SU(2)_L\times U(1)_Y$ multiplets.  
		
		\item Higher dimensional operators, obtained by integrating out heavy
		degrees of freedom, can modify the Higgs couplings. Composite Higgs
		scenario is a typical example of this type. We note that the absence
		of any signature of new physics at the LHC, till date, strongly
		motivates the use of model independent effective field theoretic
		frameworks, involving only the SM particles as the low energy
		degrees of freedom. Among the various effective theory frameworks,
		Standard Model effective field theory
		\cite{Buchmuller:1985jz,Hagiwara:1993ck,Grzadkowski:2010es,Ellis:2014dva,Jana:2017hqg,Brivio:2017vri,Ellis:2018gqa}
		and strongly interacting light Higgs scenario
		\cite{Giudice:2007fh,Contino:2013kra,Buchalla:2014eca,Chala:2017sjk}
		are worth mentioning.
	\end{itemize}
	In the present analysis, we use a simple model independent
	phenomenological Lagrangian, in the broken phase of electroweak
	symmetry, which captures the modifications of the Higgs couplings
	arising from both the above sources
	\cite{Falkowski:2013dza,Buchalla:2015qju,Cepeda:2019klc}. We expand
	the terms in the Lagrangian in powers of $h$ as well as in the number
	of derivatives. Since our primary interest lies in the production and
	decay of a single Higgs boson, we will only keep terms up to a single
	insertion of $h$. The Lagrangian involving the SM fields after the
	electroweak symmetry breaking, up to two-derivative terms is given
	below:
	\begin{equation}
	\mathcal{L}=\mathcal{L}_{(0)}+\mathcal{L}_{(2)}\,,
	\end{equation}
	where the lowest order in derivative $\mathcal{L}_{(0)}$ is given as
	\begin{align}
	\mathcal{L}_{(0)}=\frac{h}{v}\left[c_V\left(2M_W^2W^\dagger_{\mu}W^\mu+M_Z^2Z_\mu
	Z^\mu\right)-\sum_{f} c_f m_f \bar{f}f\right]\,.
	\end{align}
	The two-derivative terms, which may arise by integrating out the BSM
	states, are given by
	\begin{align}
	\mathcal{L}_{(2)}=-\frac{h}{4\pi v}\left[\alpha_e
	c_{\gamma\gamma}F_{\mu\nu}F^{\mu\nu}+\alpha_e
	c_{Z\gamma}Z_{\mu\nu}F^{\mu\nu} -\frac{\alpha_s}{2}
	c_{gg}G^a_{\mu\nu}G^{a\mu\nu}\right]\,.
	\end{align}
	The coefficients $c_i$ are free parameters capturing the impact of BSM
	physics, and to be constrained by the experimental data.  In the SM,
	$c_V = c_f = 1$ and $c_{\gamma\gamma} = c_{Z\gamma} = c_{gg} = 0$. We
	also assume those coefficients to be real, i.e. we assume the 125 GeV
	Higgs boson to be CP even. Implications of CP odd Higgs couplings have
	been discussed in
	\cite{Kobakhidze:2016mfx,Fuchs:2019ore,Fuchs:2020uoc}. The Higgs
	production cross sections and decay widths, normalized to their SM
	values, can be expressed solely in terms of these
	coefficients. Throughout this paper, we fix $c_{Z\gamma}=0$, since the
	$h\to Z\gamma$ data is too constrained from the electroweak precision
	observables and, not unexpectedly, is still unobserved at the LHC
	\cite{Aaboud:2017uhw,Sirunyan:2018tbk}. We will also assume that
	$c_\tau=c_b$ and $c_c=c_s=c_t$, to simplify the analysis.
	
	%%%%%%%%%%%%%%%%%%%%%%%%%%%%%%%%%%%%%%%%%%%%%%%%%%%%%%%%%%%%%%%%%%%%%%%%%%%%%%%%%%%%%%%%%%%%%%%%%%%%%%%%%%%%
	
	\section{Analyzing the LHC data}
	\label{LHC_data}
	
	We employ a $\chi^2$-function as defined below to constrain the
	coefficients $c_i$ using the LHC data:    
	\begin{equation}
	\chi^2=\sum_{ij}\left(\mathcal{O}^i_{\rm
		th}(\vec{c})-\mathcal{O}^i_{\rm
		exp}\right)\left[\mathcal{C}^{-1}\right]_{ij}\left(\mathcal{O}^j_{\rm
		th}(\vec{c})-\mathcal{O}^j_{\rm exp}\right)\,.
	\end{equation}
	Here $\mathcal{O}^i_{\rm exp}$ denotes the experimentally measured
	value of an observable, while $\mathcal{O}^i_{\rm th}(\vec{c})$ is the
	model prediction dependent on the parameters $c_i$. The covariance
	matrix $\mathcal{C}$ captures the experimental uncertainties and
	correlations among the different observables. For our purpose, we use
	the individual Higgs signal strength observables ($\mu$), which for a
	specific process $i\to h\to f$ is conventionally defined as
	\begin{equation}
	\mu_{i}^{f}=\frac{\sigma_i}{\sigma_i^{\rm \text{\tiny
				SM}}}\frac{B_f}{B_f^{\rm \text{\tiny
				SM}}}=\frac{\sigma_i}{\sigma_i^{\rm \text{\tiny
				SM}}}\frac{\Gamma_f}{\Gamma_f^{\rm \text{\tiny
				SM}}}\frac{\Gamma_h^{\rm \text{\tiny SM}}}{\Gamma_h}\,,
	\end{equation}
	where $\sigma_i$, $\Gamma_f$ and $B_f$ denote the cross section of the
	$i^{\rm th}$ production mode of the Higgs boson, the partial decay
	width of the Higgs into a final state $f$, and the corresponding
	branching ratio, respectively.  In the total decay width of the Higgs,
	$\Gamma_h$, we shall generally assume that the Higgs can decay only to
	the SM particles. Towards the end, however, we shall comment on the
	possibility of the Higgs boson having a non-vanishing branching
	fraction to invisible decay modes. In terms of the
	`$\kappa$-framework' \cite{Heinemeyer:2013tqa,deFlorian:2016spz}, we
	can express the cross-sections and decay widths normalized to their SM
	values as
	\begin{equation}
	\frac{\sigma_i}{\sigma_i^{\rm \text{\tiny SM}}}=\kappa_i^2,\qquad
	\frac{\Gamma_f}{\Gamma_f^{\rm \text{\tiny SM}}}=\kappa_f^2\,.
	\end{equation}
	The mapping between the $\kappa$-framework and the coefficients $c_i$
	can be found in \cite{deBlas:2018tjm}. We minimize the
	$\chi^2$-function to find the best-fit points and draw contours
	corresponding to $\Delta\chi^2=\chi^2-\chi^2_{\rm min}=2.3\,(5.99)$
	denoting regions allowed by 68\% (95\%) CL in the two dimensional
	parameter space. We also define new observables by normalizing all the
	signal strengths by that of the gold-plated $gg\to h\to ZZ^*$ process,
	measured with maximum precision.  This way the inherent uncertainties
	in the total decay width of the Higgs coming from possible invisible
	modes get eliminated.  The other advantage is that, if we assume only
	the SM particles are running inside the loops for processes like
	$gg\to h$ and $h\to \gamma\gamma$, all the ratios can be expressed in
	terms of only two variables, \textit{viz.} $c_t/c_V$ and
	$c_b/c_V$. Then the constraints from the Higgs signal strength
	measurements can be represented in a two-dimensional $c_b/c_V-c_t/c_V$
	plane. Admittedly, even if the $\Gamma_h$ dependence is eliminated in
	this approach, the errors and correlations among the ratios of signal
	strengths get slightly jacked up compared to the approach where we
	have analyzed individual signal strengths. 
	
	In this paper we primarily work with the LHC Higgs data available till
	date. In particular, we use ATLAS Run 2 data with 80 fb$^{-1}$
	luminosity \cite{Aad:2019mbh} and CMS Run 2 data with 137 fb$^{-1}$
	luminosity \cite{CMS:2020gsy}. For the purpose of comparison, we also
	show the results obtained from the combined ATLAS and CMS Run 1 data
	\cite{Khachatryan:2016vau}. As for the HL-LHC projections, with
	luminosity 3000 fb$^{-1}$, we use the SM predictions as central
	values, and the uncertainties expected to be achieved at the end of
	the HL-LHC program as  reported in \cite{Cepeda:2019klc}.  Finally we
	make some estimates of the expected precision of the Higgs coupling
	measurements using the reported sensitivities at the ILC
	($\sqrt{s}=250$ GeV, $\mathcal{L}=1.2\,\rm ab^{-1}$
	\cite{Asner:2013psa}), CEPC ($\sqrt{s}=250$ GeV, $\mathcal{L}=5.6\,\rm
	ab^{-1}$, \cite{An:2018dwb}) and FCC-ee ($\sqrt{s}=240$ GeV,
	$\mathcal{L}=5\,\rm ab^{-1}$, \cite{Abada:2019lih}).  For the clarity
	of presentation, we compiled the data used in this analysis in
	Table~\ref{data} of Appendix~\ref{app_data}. Some crucial observations
	regarding  the present data are the following. First, the processes
	involving $t\bar{t}h$ production mode have been measured with
	unprecedented precision at Run 2. Similarly, the errors for the
	$hb\bar{b}$ decay channels have got significantly reduced, in
	particular in the associated  Higgs production channel.  Besides,
	$gg\to h\to \gamma\gamma$ and  $gg\to h\to ZZ^*$ processes, which were
	already measured with less than  30\% errors in the Run 1 phase, now
	stand better with around 15\% errors  after the Run 2 data were
	analyzed. While combining ATLAS and CMS Run 2 data, we assume them to
	be independent of each other and give equal weightage to both the
	datasets in the $\chi^2$-function. However, the correlations between
	the various signal strengths, given by individual collaborations are
	included in the fit through the matrix $\mathcal{C}$.  
	
	%%%%%%%%%%%%%%%%%%%%%%%%%%%%%%%%%%%%%%%%%%%%%%%%%%%%%%%%%%%%%%%%%%%%%%%%%%%%%%%%%%%%%%%%%%%%%%%%%%%%%%%%%%%%
	
	\section{Results}
	\label{results}
	
	It has been shown in
	\cite{Falkowski:2013dza,Ciuchini:2013pca,deBlas:2016ojx} that the LEP
	data admit around $10\%-20\%$ deviation in $c_V$ from its SM value at
	95\% CL.  In the present analysis we have observed that the current
	Higgs signal strength data provide competitive, if not better, limits
	on $c_V$.

\begin{figure}[t]
		\centering
		\begin{subfigure}[t]{0.389\textwidth}
			\centering \includegraphics[trim = 0mm 0.85mm 0mm 2.6mm,
			clip,width=\linewidth]{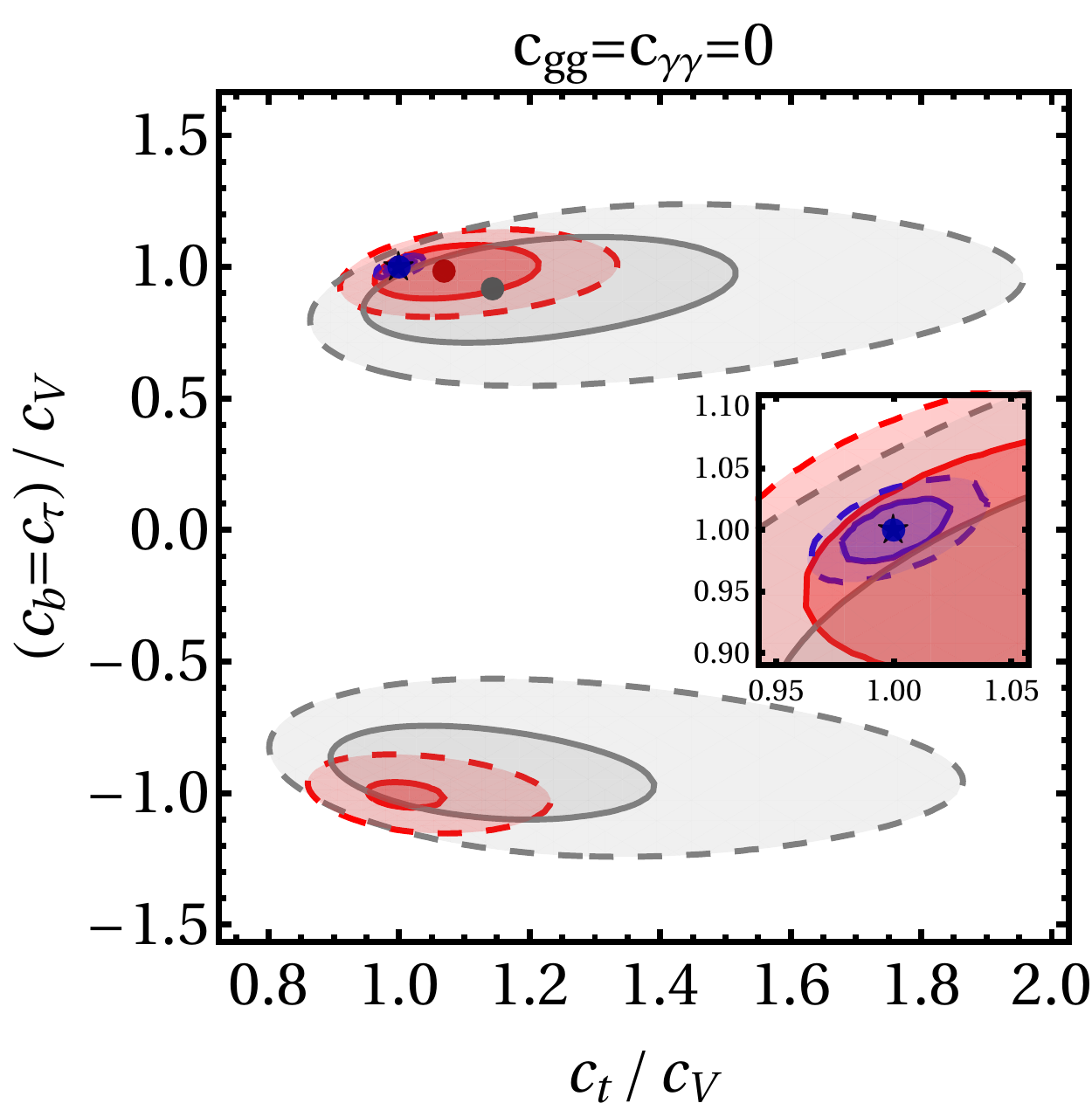}
			%\caption{}
			%\label{nmchm_s3:f1b}
		\end{subfigure}
		~
		\begin{subfigure}[t]{0.383\textwidth}
			\centering \includegraphics[trim = 0mm 0mm 0mm 0mm,
			clip,width=\linewidth]{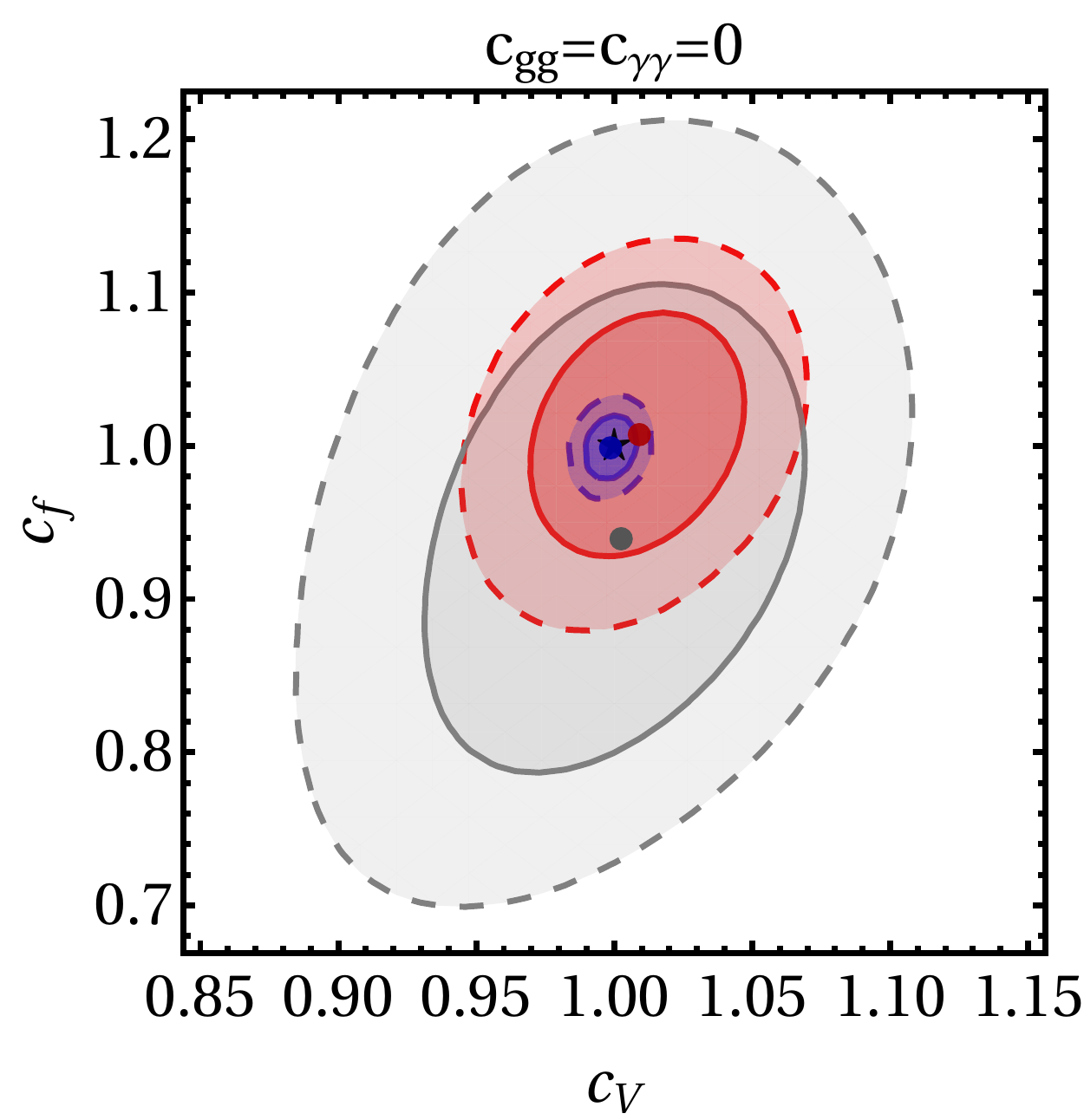}
		\end{subfigure}%
		~
		\begin{subfigure}[t]{0.18\textwidth}
			\centering \includegraphics[trim = 0mm 4mm 0mm 5mm,
			clip,width=\linewidth]{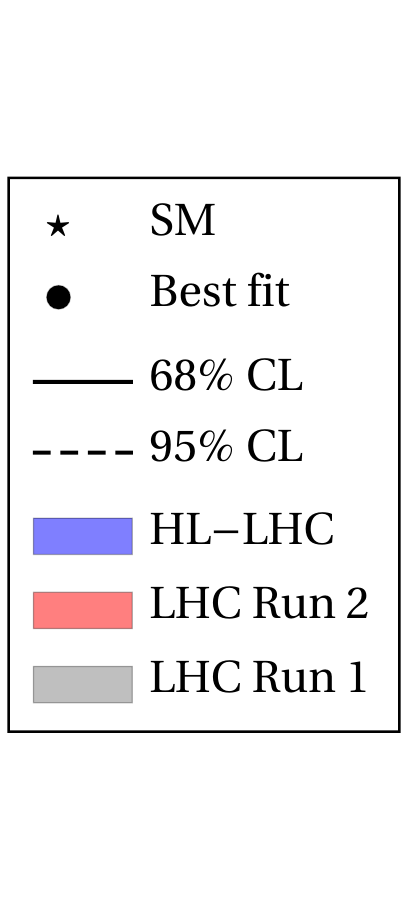}
		\end{subfigure}%
		\caption{\small\it In the left panel, the allowed regions in the plane
			of $c_b/c_V$ and $c_t/c_V$ are shown at 68\% CL (area inside the
			solid lines) and 95\% CL (area inside the dashed lines). We use the
			ratios of signal strengths from the Run 1 (grey) and Run 2 (red) LHC
			data, as well as the HL-LHC projections (blue), to extract the
			limits. The HL-LHC projection is magnified and shown in the
			inset. In the right panel, we use the individual signal strengths
			and put limits on $c_t=c_b=c_\tau=c_f$ and $c_V$. While plotting,
			$c_{gg}=c_{\gamma\gamma}=0$ is assumed.}
		\label{data_1}
	\end{figure}

	The parameter $c_t$ receives major constraints from the gluon fusion
	and $t\bar{t}h$ production modes of the Higgs boson as well as from
	its diphoton decay channel. On the other hand, constraints on $c_b$
	primarily arise from the $h\to b\bar{b}$ decay (58\% branching
	ratio). Moreover, since we have assumed $c_b=c_\tau$ in our analysis,
	data from the $h\to\tau^+\tau^-$ channel also contribute to the limits
	on $c_b$. We show in the left panel of Fig.~\ref{data_1} the allowed
	region in the $c_b/c_V-c_t/c_V$ plane, obtained using the ratios of
	the signal strengths ($\mu_i^f /\mu_{gg}^{ZZ^*}$).  The clear
	improvement from Run 1 to Run 2 is a direct consequence of more
	precise measurements of $ht\bar{t}$ and $hb\bar{b}$ couplings.
	%%%%%%%%%%%%%%%%%%%%%%%%%FIGURE%%%%%%%%%%%%%%%%%%%%%%%%%%%%%%%%%%%%%%%%%%%%%%%%%%%%%%%%%%%%%%%%%%%%%%%%%%%%%
	
	\begin{figure}[t]
		\begin{center}
			%\centering
			\includegraphics[trim = 2.5mm 0mm 0mm 0mm,
			clip,width=0.45\linewidth]{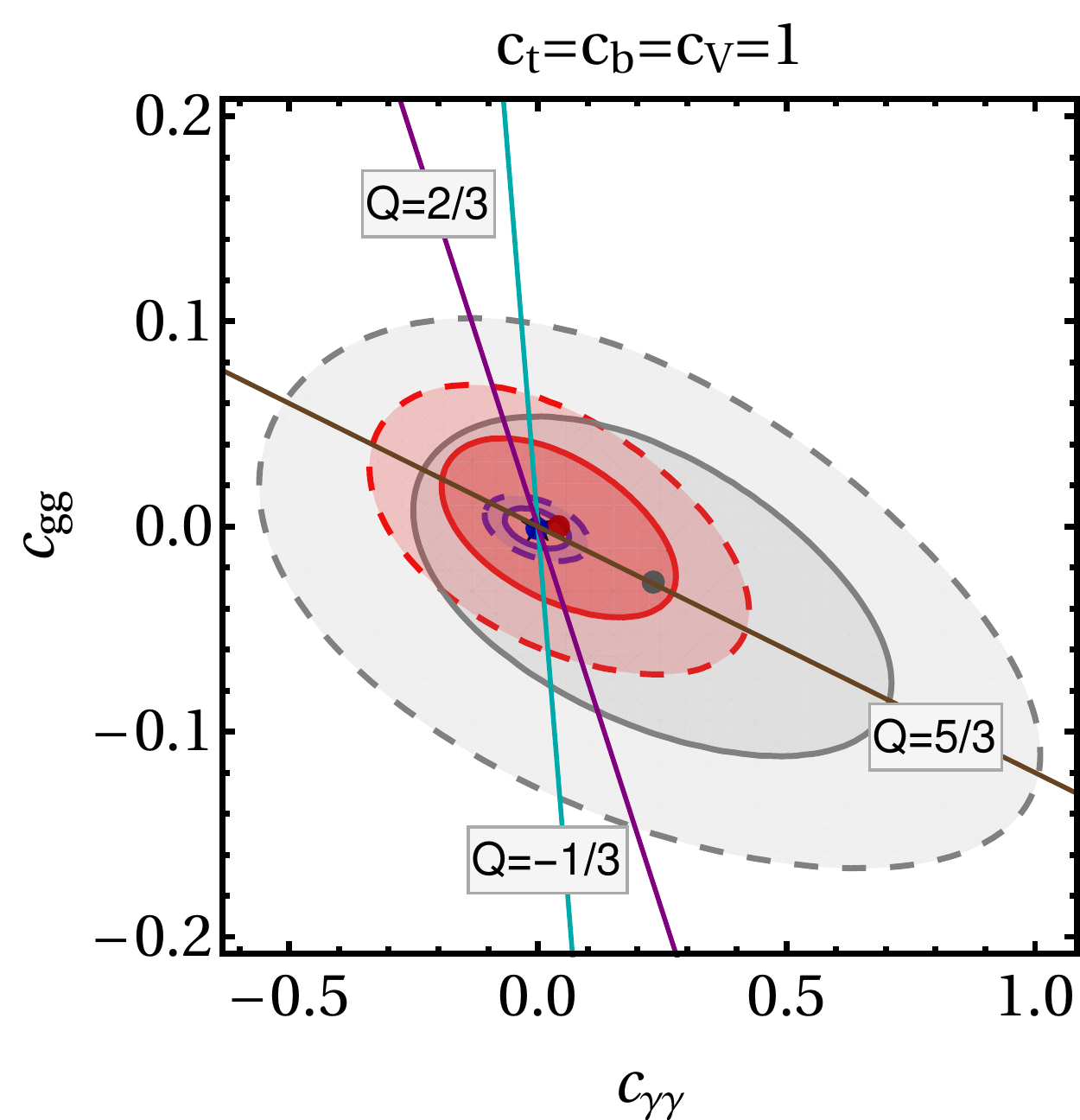}
			\caption[]{\small\it Constraints on
				$c_{gg}-c_{\gamma\gamma}$ plane at 95\% CL are
				displayed. The solid cyan, purple and brown lines
				denote the contributions to $ggh$ and
				$h\gamma\gamma$ triangle loops by color triplet BSM
				particles with electric charges $Q=-1/3$, $Q=2/3$
				and $Q=5/3$, respectively. We have fixed
				$c_f=c_V=1$. The color codes are: Run 1 (grey), Run
				2 (red), HL-LHC (blue). }
			\label{loop}
		\end{center}
	\end{figure}
	%%%%%%%%%%%%%%%%%%%%%%%%%%%%%%%%%%%%%%%%%%%%%%%%%%%%%%%%%%%%%%%%%%%%%%%%%%%%%%%%%%%%%%%%%%%%%%%%%%%%%%%%%%%%
	\begin{table}[!]
		\begin{center}
			\begin{tabular}{cccccccccc}
				\hline\hline Figure & & Quantity & & Run 1 & &
				Run 2 & & HL-LHC \\ \hline
				
				\multirow{2}{*}{Fig.~\ref{data_1} left panel }
				& & $c_b/c_V$ & & [0.55 -- 1.24] & & [0.81 --
				1.14] & & [0.95 -- 1.05] \\

				& & $c_t/c_V$ & & [0.86 -- 1.96] & & [0.91 --
				1.33] & & [0.96 -- 1.05] \\\\

				\multirow{2}{*}{Fig.~\ref{data_1} right
					panel}& & $c_f$ & & [0.70 -- 1.21] & & [0.88
				-- 1.14] & & [0.96 -- 1.04] \\

				& & $c_V$ & & [0.88 -- 1.11] & & [0.95 --
				1.07] & & [0.98 -- 1.02] \\\\

				\multirow{2}{*}{Fig.~\ref{loop}} & & $c_{gg}$
				& & [-0.17 -- 0.10] & & [-0.07 -- 0.07] & &
				[-0.02 -- 0.02] \\

				& & $c_{\gamma\gamma}$ & & [-0.57 -- 1.02] & &
				[-0.34 -- 0.43] & & [-0.11 -- 0.11]
				\\ \hline\hline
			\end{tabular}
			\caption{\small\it The range of allowed values for
				different coupling modification parameters at 95\%
				CL, extracted from the Figs.~\ref{data_1} and
				\ref{loop}, are tabulated.  Though there are two
				disjoint sets of limits on $c_b$, one on positive
				and the other on negative side, as evident from the
				left panel of Fig.~\ref{data_1}, for brevity we
				display in this Table the positive side range only.
				Assuming $c_b=c_t=c_f$, the allowed 95\% CL ranges
				of $c_f/c_V$ are obtained using the ratios of signal
				strengths as: Run~1: [0.86 -- 1.22], Run~2: [0.92 --
				1.13], HL-LHC: [0.97 -- 1.03].}
			\label{range}
		\end{center}
	\end{table} 
	In the right panel of Fig.~\ref{data_1}, we use the conventional
	approach of using the individual signal strengths to extract the
	limits. Here we assume $c_t=c_b=c_\tau=c_f$ to show the allowed region
	in the $c_f-c_V$ plane. Our results for Run 1 are consistent with
	those found in \cite{deBlas:2018tjm}, validating our fitting
	method. For the Run 2 data we find $\chi^2_{\rm min}=21.53$ and
	$\chi^2_{\rm SM}-\chi^2_{\rm min}<1$, indicating that the SM fits the
	data very well. In Table~\ref{range} we display the allowed ranges of
	parameters at 95\% CL.  Two major points are worth noting here. First,
	the limits on $c_V$ from the Run 2 data are already competitive to
	those obtained from the electroweak precision tests.  This happened
	primarily due to the increasingly precise measurements of the $gg\to
	h\to ZZ^*$ and $gg\to h\to WW^*$ processes. Second, the window for new
	physics through $c_f$ has significantly narrowed down, only
	$10\%-15\%$ deviation is allowed from the SM reference point. This
	improvement in Yukawa force measurement helps discriminate various BSM
	scenarios.  We note that the combined Run 1 + Run 2 data improve the
	limits obtained from Run 2 data alone by at most 2\% -- 3\%. While
	future $e^+e^-$ colliders can directly measure the $hVV$ and
	$hb\bar{b}$ couplings with much higher precision, $ht\bar{t}$ coupling
	will get constrained only indirectly. We have estimated that at the
	future Higgs factories like ILC (CEPC), the uncertainties on $c_V$
	and $c_f$ would be around $0.01~(0.004)$ and $0.03~(0.009)$ at 95\%
	CL, respectively. In FCC-ee as well, the uncertainties would reduce by
	almost an order of magnitude to $\mathcal{O}(10^{-3})$ in comparison
	to HL-LHC. In obtaining the above constraints we have assumed
	$c_{gg}=c_{\gamma\gamma}=0$. The inherent assumption is that any new
	BSM particle(s) which might have contributed to the triangle loops
	creating the effective $ggh$ and $h\gamma\gamma$ vertices are
	sufficiently heavy and decoupled.
	
	Then we go to the other extreme. Keeping $c_b=c_t=c_\tau=c_V=1$, we
	display the limits in $c_{gg}-c_{\gamma\gamma}$ plane in
	Fig.~\ref{loop}.  Here we capture the effects of the new BSM particles
	floating in the triangle loops, e.g. if the SM is extended with
	additional colored and electrically charged particles. The solid lines
	represent the contributions from colored particles, transforming as
	triplets of $\rm SU(3)_c$ and having electric charges $Q=-1/3$ (cyan),
	$Q=2/3$ (purple) and $Q=5/3$ (brown), respectively. The exact location
	of a model-point on each straight line, however, depends on the mass
	and model-dependent couplings of the new particles with the Higgs
	boson \cite{Falkowski:2013dza}.
	
	If a non-vanishing branching fraction for the invisible decay mode
	($B_{\rm inv}$) of the Higgs boson is admitted, all the individual
	signal strengths of the Higgs boson receive a scaling by an overall
	factor of $\left(1-B_{\rm inv}\right)$. Assuming $c_f=c_V=1$ and
	$c_{gg}=c_{\gamma\gamma}=0$, we observe that the Run 1 (Run 2) data
	exclude $B_{\rm inv}\gtrsim18\%$ (8\%), while the HL-LHC
	(ILC/CEPC/FCC-ee) would exclude $B_{\rm inv}\gtrsim 3\%$ (1\%).
	Admittedly, these limits will relax considerably, if deviations in
	$c_i$ parameters are allowed ({\em e.g.} %keeping many parameters
	the Particle Data Group excludes a rather conservative $B_{\rm
		inv}\gtrsim 24\%$ \cite{Tanabashi:2018oca}).
	%%%%%%%%%%%%%%%%%%%%%%%%%%%%%%%%%%%%%%%%%%%%%%%%%%%%%%%%%%%%%%%%%%%%%%%%%%%%%%%%%%%%%%%%%%%%%%%%%%%%%%%%%%%% 
	\subsection{Composite Higgs models}
	
	In generic composite Higgs scenario, the modification in the $hVV$
	coupling is universal \cite{Liu:2018vel,Liu:2018qtb}
	\begin{equation}
	c_V=\sqrt{1-\xi}\,,
	\end{equation}
	where $\xi=v^2/f^2$ parametrizes the hierarchy between the electroweak
	scale and the composite scale $f$. The Yukawa couplings, however,
	depend on the details of the particular model. In the minimal
	composite Higgs model, with coset SO(5)/SO(4)
	\cite{Contino:2010rs,Panico:2015jxa,Csaki:2016kln}, the Yukawa
	coupling modifiers are controlled by the specific representations of
	SO(5) in which the SM quarks and leptons are embedded. A generic
	parametrization for $c_f$ in such cases can be given as
	\begin{equation}
	c_f=1+\Delta_f\xi\,,
	\end{equation} 
	where $\Delta_f$ is a free parameter which depends on the number of
	Yukawa operators. If only one Yukawa operator exists, as in cases
	where the SM fermions are embedded in $\bf 4$, $\bf 5$ or $\bf 10$ of
	SO(5), $c_f$ is determined only by $\xi$
	\cite{Contino:2003ve,Agashe:2004rs,Panico:2011pw,Marzocca:2012zn,Carena:2014ria,Sanz:2017tco}.
	For example, if the top quark is embedded in the fundamental $\bf 5$
	of SO(5) ($\rm MCHM_{5}$), we find $\Delta_t=-3/2$, while putting the
	top in spinorial $\bf 4$ ($\rm MCHM_{4}$), we obtain
	$\Delta_t=-1/2$. When more than one operator can be constructed,
	$\Delta_f$ depends on the microscopic parameters of the composite
	dynamics. Such possibilities may occur when either of the left- or
	right-handed fermions are embedded in the symmetric $\bf 14$
	dimensional representations of SO(5)
	\cite{Montull:2013mla,Carena:2014ria,Carmona:2014iwa,Kanemura:2016tan,Gavela:2016vte,Liu:2017dsz,Banerjee:2017wmg}.
	
\begin{figure}[t]
		\centering
		\begin{subfigure}[t]{0.385\textwidth}
			\centering \includegraphics[trim = 0mm 0.0mm 0mm 0mm,
			clip,width=\textwidth]{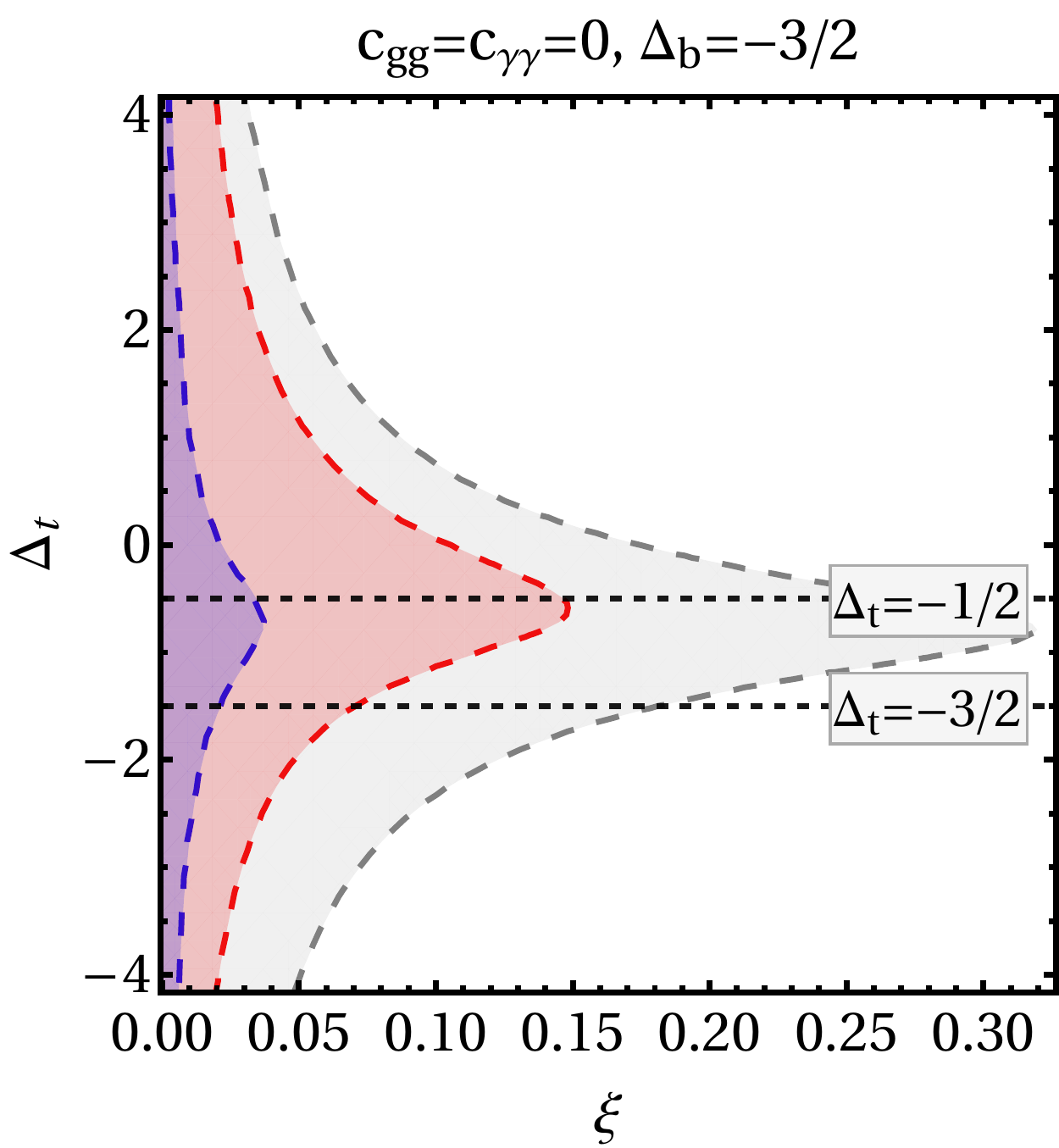}
			%\caption{}
			%\label{nmchm_s3:f1b}
		\end{subfigure}
		~
		\begin{subfigure}[t]{0.385\textwidth}
			\centering
			\includegraphics[width=\textwidth]{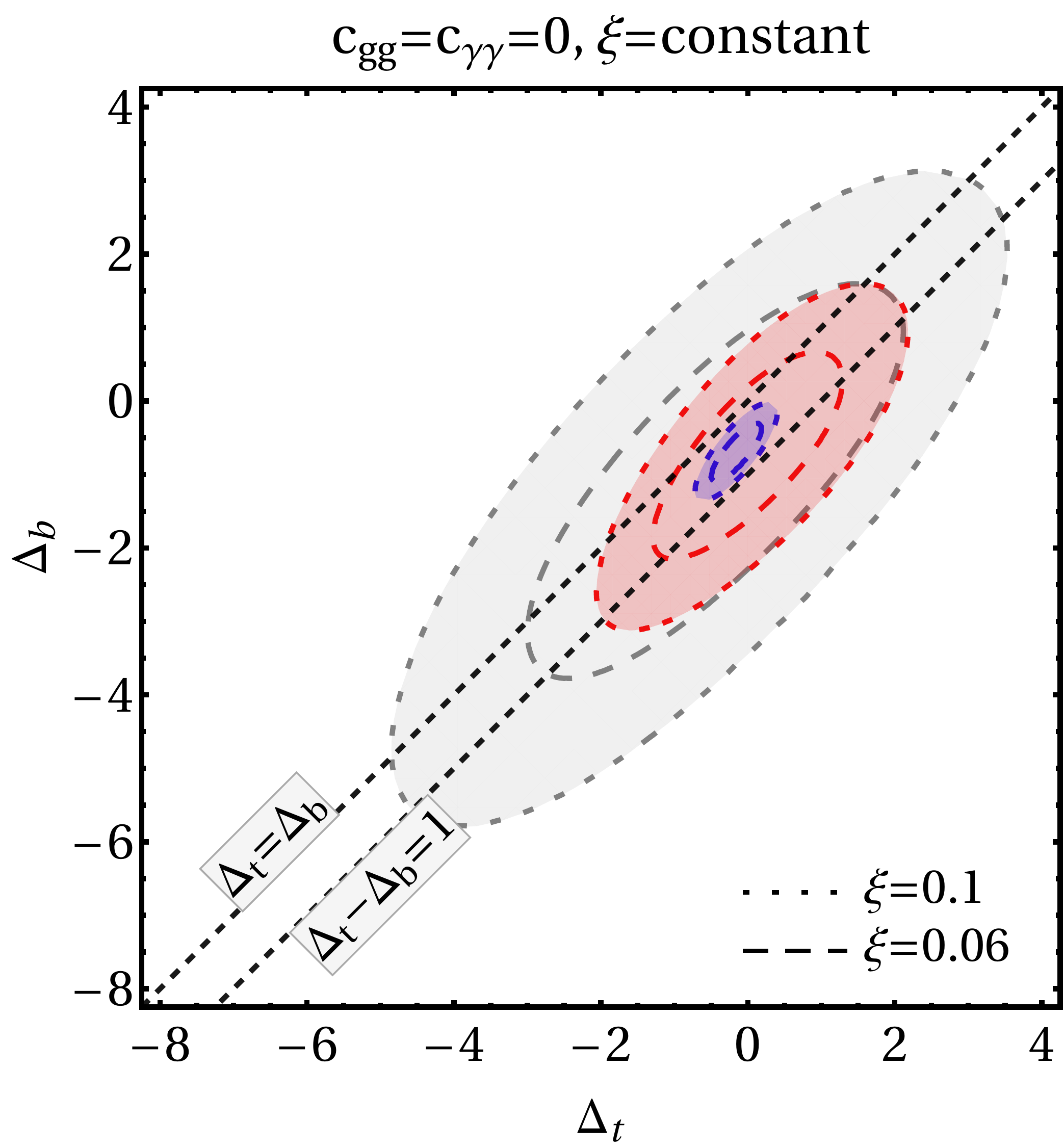}
			%\caption{}
			%\label{nmchm_s3:f1a}
		\end{subfigure}%
		\caption{\small\it The 95\% CL allowed regions for minimal
			composite Higgs models are shown. In left panel, we fixed
			$\Delta_b=-3/2$, while in the right panel, we chose
			$\xi=0.1$ (dotted) and $\xi=0.06$ (dashed). The horizontal
			black dashed lines in the left panel correspond to
			$\Delta_t=-1/2$ ($MCHM_{4}$), and $\Delta_t=-3/2$
			($MCHM_{5}$). In the right panel, similar lines represent
			the contours of $\Delta_t-\Delta_b=0,\, 1$. The color codes
			are: Run 1 (grey), Run 2 (red), HL-LHC (blue).}
		\label{chm_plot_1}
	\end{figure}

	Here we discuss three specific cases for which we have obtained new
	limits:
	\begin{itemize}
		\item $\Delta_t=\Delta_b=\Delta_\tau=-3/2$ ($\rm MCHM_{5}$): This is
		an oft-quoted example when both the left- and right-chiral top quark
		are kept in {\bf 5} of SO(5), necessarily  yielding $c_{f,V}<1$. In
		this case, the $\chi^2$- function depends  on a single parameter
		$\xi$. The constraints from Run 2 data are  slightly stronger than
		expectation as the data show a small bias  towards the $c_{f,V}>1$
		region (see right panel of Fig.~\ref{data_1}).  We obtain $f\gtrsim
		1.03$ TeV at 95\% CL using the Run 2 data, while  in HL-LHC we
		expect $f\gtrsim 1.8$ TeV. In proposed Higgs factories this limit is
		expected to be further strengthened as $f\gtrsim 2.4$ TeV (ILC),
		$f\gtrsim 3.3$ TeV (CEPC), and $f\gtrsim 3.1$ TeV (FCC-ee).
		
		\item $\Delta_b=\Delta_\tau=-3/2$: Here, we keep $\Delta_t$ as a free
		parameter, which implies either the left- or the right-handed top
		quark is embedded in $\bf 14$ of SO(5). The allowed region at 95\%
		CL in the $\Delta_t-\xi$ plane is shown in the left panel of
		Fig.~\ref{chm_plot_1}. Clearly, the constraint on $f$ gets relaxed,
		as alluded in \cite{Banerjee:2017wmg}. For a generic value of
		$\Delta_t$, we obtain the most conservative limit $f\gtrsim 650$ GeV
		after inclusion of the Run 2 data.
		
		\item $\xi={\rm constant}$: We fix two representative values of $\xi =
		0.1$ and $0.06$, to put simultaneous limits in $\Delta_b-\Delta_t$
		plane as shown in the right panel of Fig.~\ref{chm_plot_1}. We
		observe that while the present data have not yet gathered enough
		strength to discriminate between the choices of representations in
		which the top and bottom quarks are embedded, future measurements
		with better statistical significance can do the job.
	\end{itemize}
	%%%%%%%%%%%%%%%%%%%%%%%%%FIGURE%%%%%%%%%%%%%%%%%%%%%%%%%%%%%%%%%%%%%%%%%%%%%%%%%%%%%%%%%%%%%%%%%%%%%%%%%%%%%
	%%%%%%%%%%%%%%%%%%%%%%%%%%%%%%%%%%%%%%%%%%%%%%%%%%%%%%%%%%%%%%%%%%%%%%%%%%%%%%%%%%%%%%%%%%%%%%%%%%%%%%%%%%%%
	We have kept $c_{gg}=c_{\gamma\gamma}=0$. This is motivated by the
	observation that in the composite pseudo-Goldstone Higgs scenario the
	top partner loop contribution cancels against the contribution of the
	wave function renormalization of the top quark
	\cite{Azatov:2011qy,Montull:2013mla}. The current direct search limit
	on the vector-like top-partners is around $m_*\sim g_*f\gtrsim 1.5$
	TeV \cite{Aaboud:2018pii,Aaboud:2018xpj,Sirunyan:2019sza}. Considering the strong coupling $1\ll g_*<4\pi$, we clearly
	observe that the Higgs coupling measurements provide somewhat stronger limits
	on the compositeness scale. However, the limits from the electroweak
	precision observables remain comparable, \textit{albeit} with
	additional model dependence coming from incalculable UV dynamics.
	
	Composite Higgs models are often seen as dual to some variants of the
	weakly coupled warped extra dimensional models using the AdS / CFT
	correspondence \cite{Maldacena:1997re}. We take a custodial
	Randall-Sundrum (RS) setup with the Higgs boson localized near the IR
	brane to study the constraints on the scale of the Kaluza-Klein states
	($M_{\rm KK}$)
	\cite{Randall:1999ee,Agashe:2003zs,Casagrande:2010si,Malm:2013jia,Malm:2014gha}.
	Adapting the expressions for the Higgs coupling modifiers from
	\cite{Malm:2014gha}, including the Run 2 data, we obtain a
	conservative lower limit on the mass of the first excited KK-gluon,
	$M_g\gtrsim 9$ TeV (which translates into $M_{\rm KK}\gtrsim 3.7$
	TeV).  The projected limit from HL-LHC is $M_g\gtrsim 13$ TeV.
	
	%%%%%%%%%%%%%%%%%%%%%%%%%%%%%%%%%%%%%%%%%%%%%%%%%%%%%%%%%%%%%%%%%%%%%%%%%%%%%%%%%%%%%%%%%%%%%%%%%%%%%%%%%%%%
	
	\subsection{Multi-Higgs models}

	%%%%%%%%%%%%%%%%%%%%%%%%%FIGURE%%%%%%%%%%%%%%%%%%%%%%%%%%%%%%%%%%%%%%%%%%%%%%%%%%%%%%%%%%%%%%%%%%%%%%%%%%%%%
	\begin{figure}[t!]
		\centering
		\begin{subfigure}[t]{0.395\textwidth}
			\centering \includegraphics[trim = 0mm 1.0mm 0mm 0mm,
			clip,width=\linewidth]{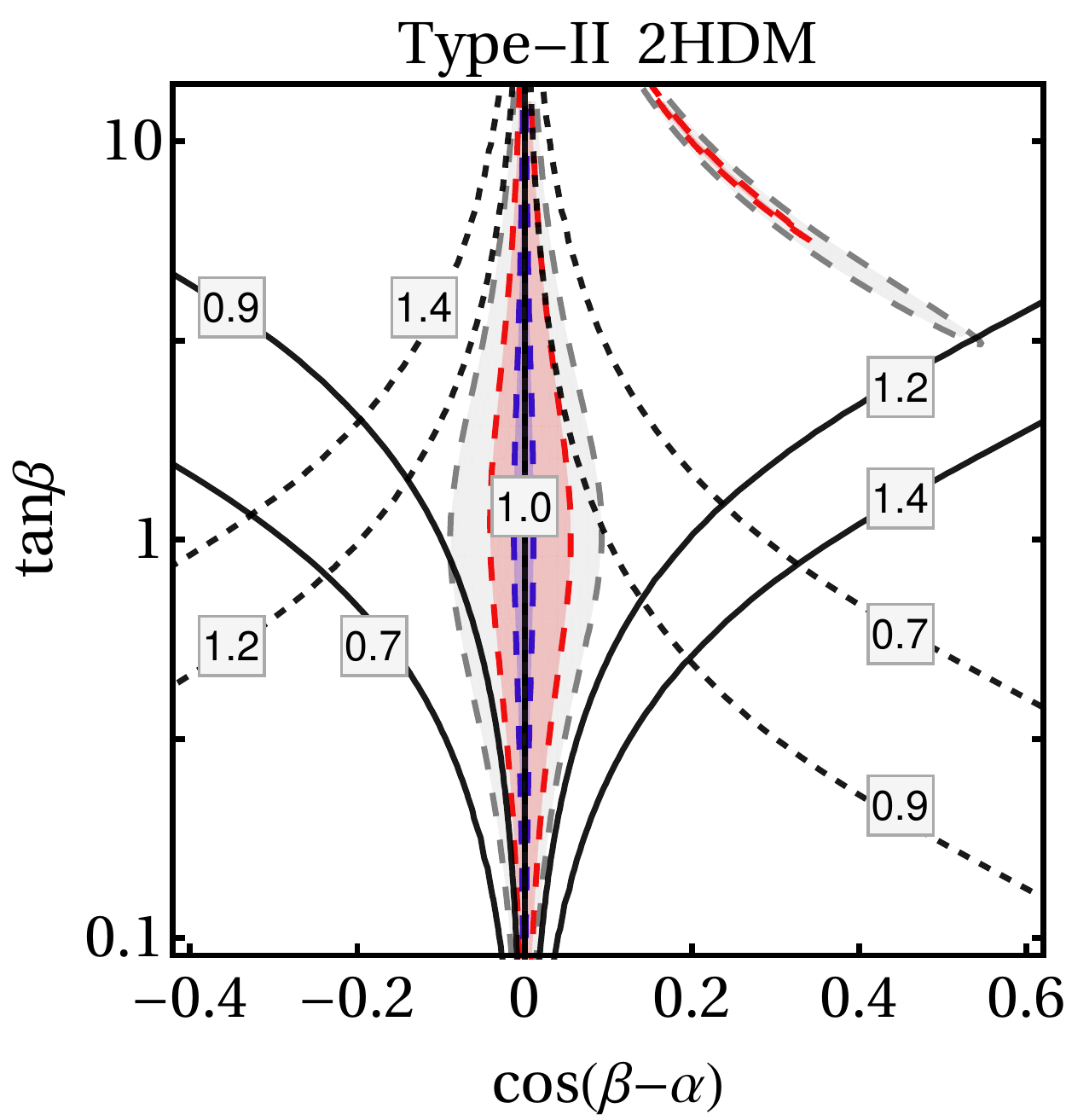}
			%\caption{}
			%\label{nmchm_s3:f1b}
		\end{subfigure}
		~
		\begin{subfigure}[t]{0.395\textwidth}
			\centering \includegraphics[trim = 0mm 1.0mm 0mm 0mm,
			clip,width=\linewidth]{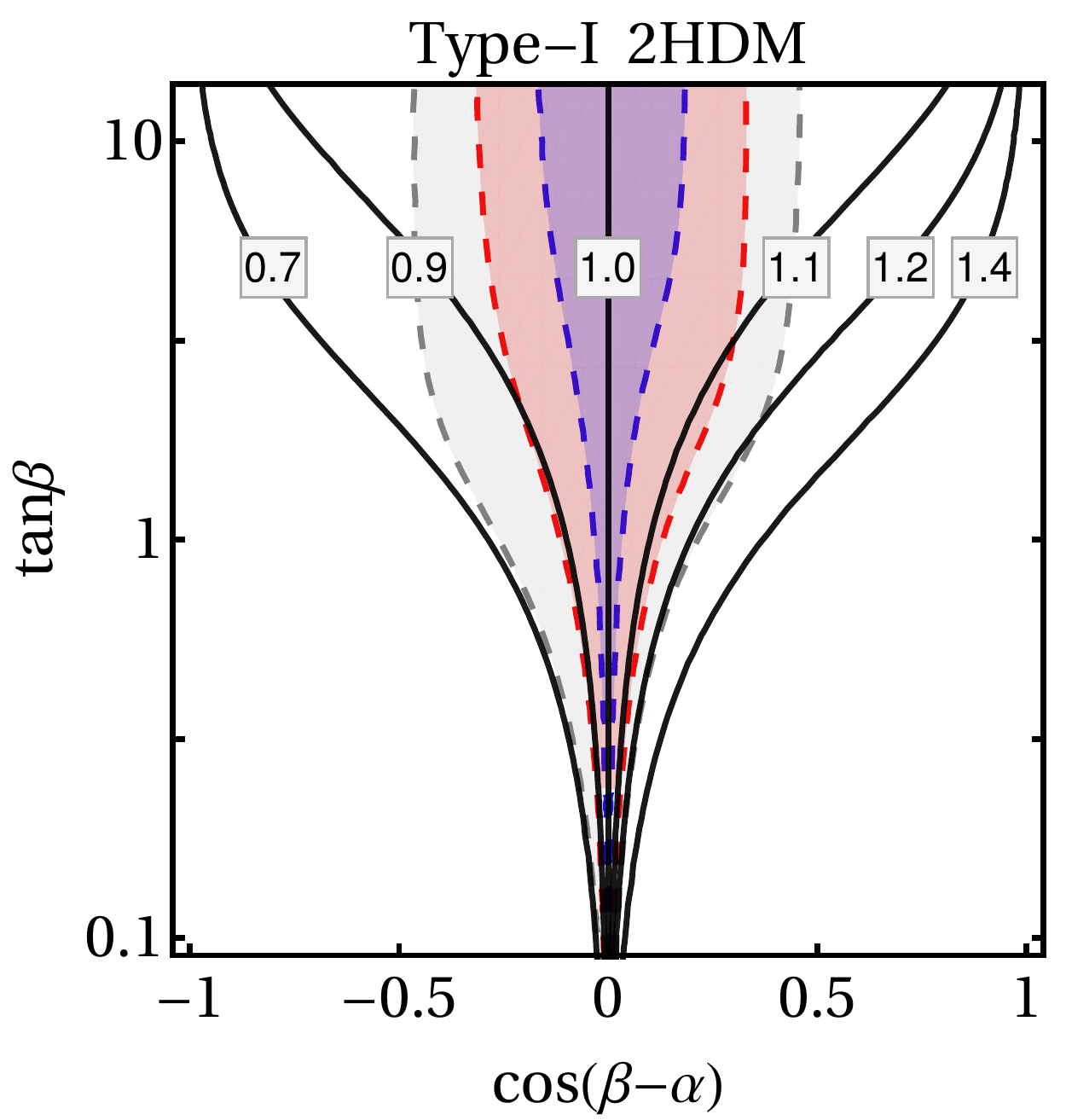}
			%\caption{}
			%\label{nmchm_s3:f1a}
		\end{subfigure}%
		~\\\vspace{0.5cm}
		\begin{subfigure}[t]{0.405\textwidth}
			\centering \includegraphics[trim = 0mm 1.0mm 0mm 0mm,
			clip,width=\linewidth]{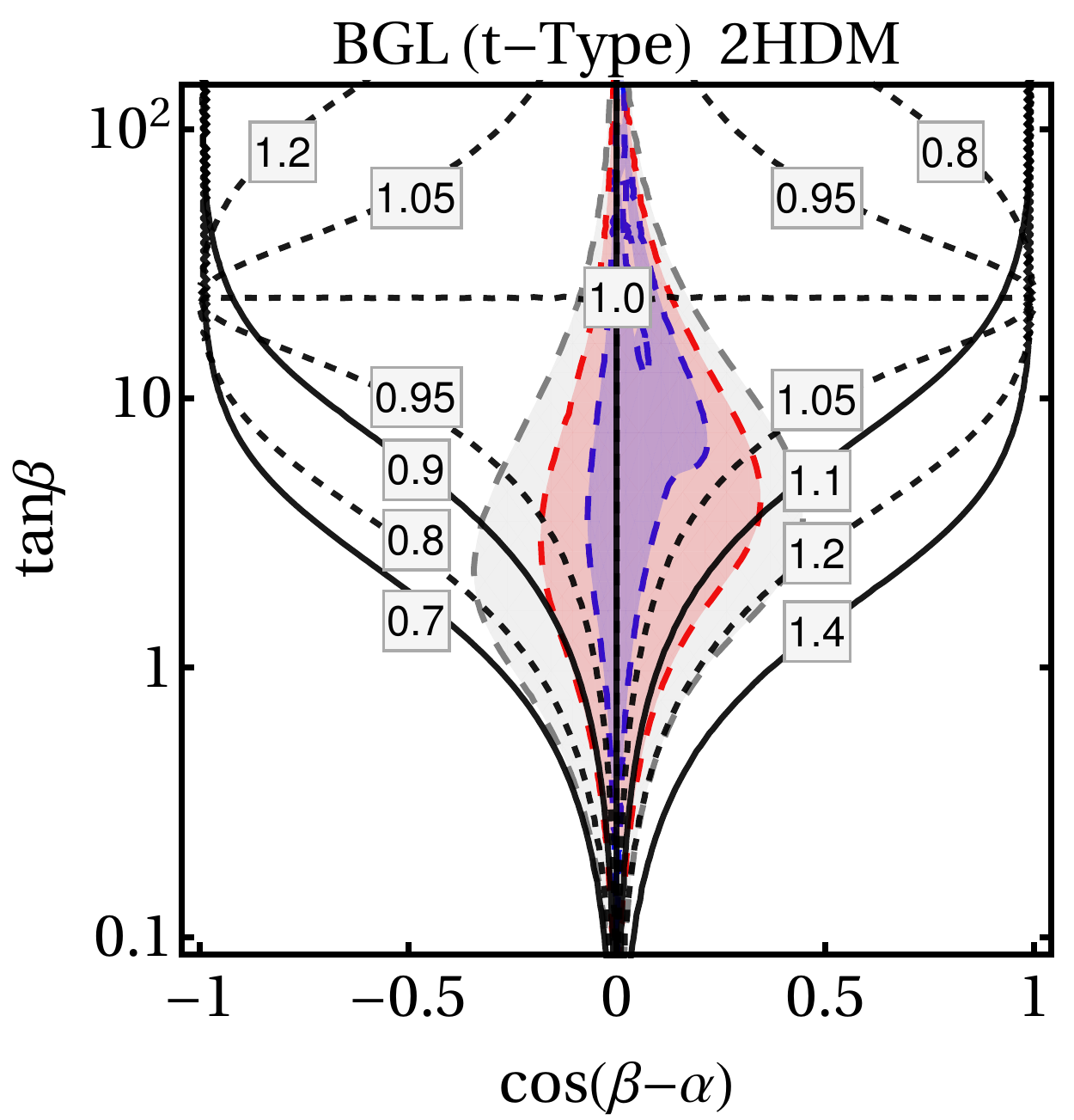}
			%\caption{}
			%\label{nmchm_s3:f1a}
		\end{subfigure}%
		~~
		\begin{subfigure}[t]{0.395\textwidth}
			\centering \includegraphics[trim = 0mm 0mm 0mm 0mm,
			clip,width=\linewidth]{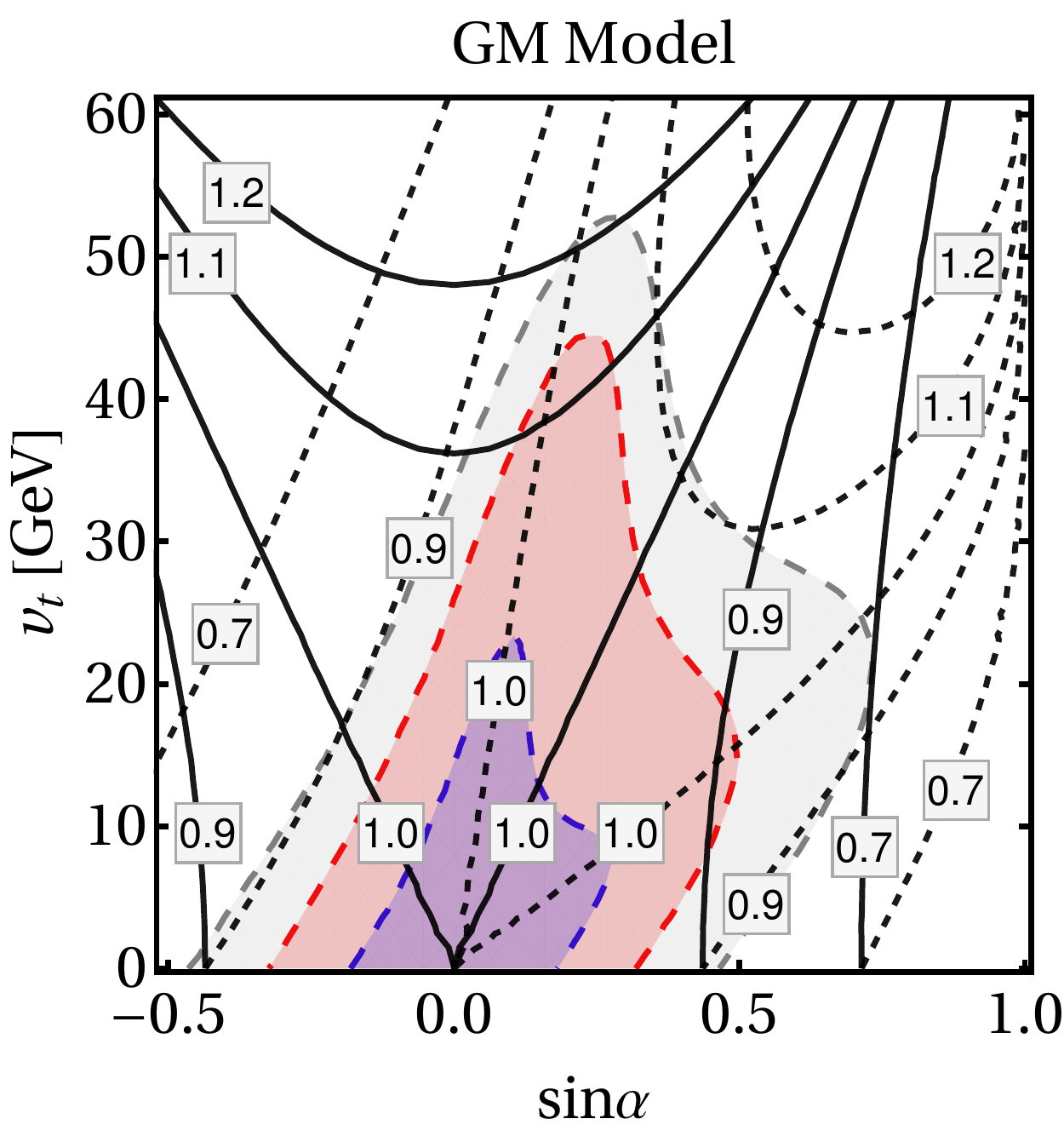}
			%\caption{}
			%\label{nmchm_s3:f1a}
		\end{subfigure}%
		\caption{\small\it The top-left (top-right) panels show the
			constraints on parameter space of Type-II (Type-I) 2HDM
			respectively, while the bottom-left panel corresponds to the
			BGL (t-type) model. For 2HDM, the solid (dashed) black lines
			denote the contours of constant $c_t/c_V$ ($c_b/c_V$). The
			results for Type-I and Type-II 2HDM conform to those
			obtained in \cite{Sirunyan:2018koj,Aad:2019mbh}. The
			bottom-right panel displays the limits on the GM model, for
			which the solid lines denote the contours for $c_t=c_b$ and
			the dashed lines denote the same for $c_V$. The color codes
			are: Run 1 (grey), Run 2 (red), HL-LHC (blue).}
		\label{2hdm}
	\end{figure}
	%%%%%%%%%%%%%%%%%%%%%%%%%%%%%%%%%%%%%%%%%%%%%%%%%%%%%%%%%%%%%%%%%%%%%%%%%%%%%%%%%%%%%%%%%%%%%%%%%%%%%%%%%%%%
	Here we deal with theories involving multiple Higgs bosons with
	non-trivial ${\rm SU(2)_L}\times {\rm U(1)_Y}$ charges. The question
	is whether the 125 GeV Higgs boson discovered at the LHC is the only
	one of its genre. Since a long time, searches for additional Higgs
	multiplets are going on in colliders including the LHC.  The most
	trivial extension of the SM is the addition of a gauge singlet CP-even
	scalar boson
	\cite{Robens:2015gla,Robens:2016xkb,Lewis:2017dme,Adhikari:2020vqo}. Due
	to the ensuing doublet-singlet scalar mixing, parametrized by an angle
	$\alpha$, the 125 GeV Higgs couplings pick up a factor of
	$\cos\alpha$. At 95\% CL, from Run 1 (Run 2) data we obtain
	$\sin\alpha\lesssim 0.43~(0.28)$, while the HL-LHC expectation is
	$\sin\alpha\lesssim 0.17$. The estimated limits in case of future
	Higgs factories are $\sin\alpha\lesssim 0.11$ for ILC, while
	$\sin\alpha\lesssim 0.09$ for CEPC and FCC-ee.
	
	Now we focus on two-Higgs doublet models (2HDM)
	\cite{Branco:2011iw,Bhattacharyya:2013rya,Craig:2013hca,Eberhardt:2013uba,Belanger:2013xza,Dumont:2014wha,Bhattacharyya:2015nca,Chowdhury:2017aav,Ren:2017jbg,Lu:2015qqa}. The
	$hVV$ coupling modifications in 2HDM depend on two mixing angles as
	\begin{equation}
	c_V=\sin(\beta-\alpha)\,.
	\end{equation}
	Above, the angle $\beta$ parametrizes the mixing between the two
	doublets, while $\alpha$ is a measure of mass-mixing between the two
	CP even neutral scalars. In Type-II 2HDM, which also forms the basis
	of constructing the minimal supersymmetric standard model, the Yukawa
	coupling modifiers are given by
	\begin{equation}
	c_t=\frac{\cos\alpha}{\sin\beta}, \qquad
	c_b=-\frac{\sin\alpha}{\cos\beta}\,.
	\end{equation}
	Note that, $c_t\ne c_b$ in this case. We have shown the limits on
	$\tan\beta$ and $\cos(\beta-\alpha)$ in the top-left panel of
	Fig.~\ref{2hdm}. The narrow window of allowed region around the
	alignment limit $\beta-\alpha=\pi/2$ has shrunk considerably with
	respect to earlier data. In Type-I 2HDM, however, the top and bottom
	Yukawa couplings are modified by the same factor as
	\begin{equation}
	c_t=c_b=\frac{\cos\alpha}{\sin\beta}\,.
	\end{equation}
	In this case, constraints are displayed in the top-right panel of
	Fig.~\ref{2hdm}. The results we found for both Type-I and Type-II 2HDM
	are compatible with those reported in
	\cite{Sirunyan:2018koj,Aad:2019mbh}. The direct searches of heavy
	Higgs ($H$) and pseudoscalar Higgs ($A$) bosons furnish a
	complementary tool to constrain the 2HDM parameter space. Around the
	alignment limit, which seems to be strongly favored by the Higgs data,
	the sensitivity of the direct search limits is poor
	\cite{Chowdhury:2017aav}. Away from the alignment zone and for a heavy
	Higgs mass $m_H\lesssim1$ TeV, $H\to hh$ (above the di-Higgs
	threshold) and $H\to VV$, $H\to \tau^+\tau^-$ channels provide
	dominant constraints. The limits for the pseudoscalar Higgs search are
	important for $m_A\lesssim 1$ TeV, similar to that of heavy Higgs. A
	special category of 2HDM postulated by Branco, Grimus and Lavoura (the
	BGL scenario) admits flavor changing neutral current interactions at
	the tree level, suppressed by the elements of the
	Cabibbo-Kobayashi-Maskawa (CKM) matrix
	\cite{Branco:1996bq,Bhattacharyya:2014nja,Botella:2015hoa}. In some
	variants of the BGL model (t-type), the expression for $c_t$ resembles
	that of Type-I 2HDM, while $c_b$ receives an additional contribution
	proportional to $\left(\tan\beta+\cot\beta\right)$ as follows
	\cite{Bhattacharyya:2014nja}:
	\begin{equation}
	\label{BGL}
	c_t=\frac{\cos\alpha}{\sin\beta}\,, \quad
	c_b=\frac{\cos\alpha}{\sin\beta}-\cos\left(\beta-\alpha\right)
	\left(\tan\beta + \cot\beta\right) \left(1-|V_{tb}|^2\right)\,.
	\end{equation}
	In the low $\tan\beta\lesssim1$ regime, the constraints on the BGL
	(t-type) model follow that of the Type-I scenario (see the bottom-left
	panel of Fig.~\ref{2hdm}). But with the increasing $\tan\beta\gg 10$,
	owing to the second term in $c_b$ in Eq.~\eqref{BGL}, tighter limits
	are obtained compared to the Type-I model. Notably, the LHC data
	provide complementary constraints in the low $\tan\beta$ region, which
	is otherwise less sensitive to the flavor observables
	\cite{Bhattacharyya:2014nja}.
	
	Next, we discuss the triplet-extended scenarios, in particular
	Georgi-Machacek (GM) model
	\cite{Georgi:1985nv,Chanowitz:1985ug,Gunion:1989ci,Belanger:2013xza,Hartling:2014zca,Hartling:2014aga,Chiang:2014bia,Degrande:2017naf,Chiang:2018cgb,Ghosh:2019qie}. In
	this model the custodial symmetry is protected by the tree level
	scalar potential, even if the triplets receive a vev ($v_t$).  Without
	going into the details of the model, we give the expressions for
	$c_V$, $c_t$ and $c_b$ for this case as
	\begin{equation}
	c_V=\cos\alpha\cos\beta+2\sqrt{\frac{2}{3}}\sin\alpha\sin\beta\,,\qquad
	c_t=c_b=\frac{\cos\alpha}{\cos\beta}\,. 
	\end{equation}
	The limits obtained using the Run 1 and Run 2 data from the LHC and
	the HL-LHC projections are shown in the $v_t-\sin\alpha$ plane in the
	bottom-right panel of Fig.~\ref{2hdm}. In analogy to the 2HDM
	scenario, here also the Higgs data give stronger constraints around
	$\sin\alpha\sim0$ in comparison to the limits coming from the direct
	searches of the heavy states \cite{Chiang:2018cgb}.
	
	A few comments on our analysis are in order. First, in deriving these
	constraints for 2HDM and the GM model, we assumed that the
	contribution of the charged Higgs bosons decouple in the
	$h\gamma\gamma$ decay width and thus can be neglected. However, as
	shown in \cite{Bhattacharyya:2014oka,Das:2018vkv} the decoupling of
	the charged Higgs contribution to diphoton decay channel depends on
	the details of the particular model in question. Indeed, our limits
	would change accordingly. Second, the limits are obtained assuming
	only renormalizable interactions. The presence of higher dimensional
	operators
	\cite{Chala:2017sjk,DiazCruz:2001tn,Kikuta:2011ew,Crivellin:2016ihg,Karmakar:2017yek,Banerjee:2017wmg,Chala:2018opy,Banerjee:2019gmr}
	would lead to further modifications of all the couplings in addition
	to what comes out of the mixing in the renormalizable setup. As shown
	in \cite{Karmakar:2018scg} in the context of 2HDM and in
	\cite{Banerjee:2019gmr} for the GM model, these additional
	modifications, seeping through extra coefficients, would leave
	indelible  imprint on the ranges of the model parameters.
	
	%%%%%%%%%%%%%%%%%%%%%%%%%%%%%%%%%%%%%%%%%%%%%%%%%%%%%%%%%%%%%%%%%%%%%%%%%%%%%%%%%%%%%%%%%%%%%%%%%%%%%%%%%%%
	
	\section{Conclusions and outlook}
	\label{conclusion}
	
	We summarize below the important points raised in this paper. The LHC
	Run 2 data contain a significantly improved information on the Yukawa
	couplings. Their inclusion has allowed us to extract important
	limits. 
	\begin{itemize}
		
		\item The ATLAS and CMS Collaborations have made an important
		breakthrough in getting a grip on the Yukawa force for the first
		time.  If the Higgs boson turns out to be elementary, then it
		signifies the observation of a new fundamental force. The
		experimental measurements have made a huge impact in constraining
		the allowed region of BSM physics manifesting through modified
		Yukawa couplings. The Run 2 data are particularly instrumental in
		squeezing the 2$\sigma$ BSM window in the Yukawa couplings from 25\%
		to 15\% around their SM values when compared to the performance of
		the Run 1 data. HL-LHC would bring it down to within 5\%.  The
		limits on $hVV$ ($V=W,Z$) couplings from the LHC are now competitive
		with those obtained from electroweak precision tests. The Run 1 (Run
		2) data allow not more than 18\% (8\%) of the total branching
		fraction of the Higgs boson in the invisible channel. However,
		larger leak into invisible mode can be accommodated if the $hVV$ and
		$hf\bar{f}$ couplings substantially deviate from their SM reference
		points.
		
		\item We consider a few motivated BSM scenarios and recast the
		constraints from our model independent analysis on the parameter
		space of those specific models using the latest Higgs data. We have
		observed that, in the context of the SO(5)/SO(4) minimal composite
		Higgs model, more precise measurements of Yukawa forces have
		improved the limits on the compositeness scale. The limits depend on
		the representations of SO(5) in which we embed the left- and
		right-chiral top quark. At 95\% CL, our new limits are
		$$ f \gtrsim 650 ~{\rm GeV ~(most ~conservative)}\,, \qquad f
		\gtrsim 1.03 ~{\rm TeV} ~({\rm MCHM_{5}}).
		$$
		We have shown how the future HL-LHC data would further sharpen the
		limits. In the RS scenarios with the Higgs boson localized near the
		IR brane, the first excited KK-gluon weighs more than
		$\mathcal{O}(10)$ TeV. The exact limit depends on the details of the
		model parameters.
		
		\item The amount of mixing between the SM Higgs with any additional
		scalar singlet is observed to be rather constrained by the present
		data, given by $\sin\alpha\lesssim 0.28$. For Type-II 2HDM, only a
		narrow region around the alignment limit is acceptable, while for
		the Type-I case a considerable area in the large $\tan\beta$ region
		is still allowed. In the BGL (t-type) model the constraints in the
		low $\tan\beta\lesssim 1$ region are in the same ballpark as in the
		Type-I scenario, while for $\tan\beta\gg 10$ the BGL (t-type)
		receives stronger constraints than Type-I.  We have also shown that
		for the Georgi-Machacek model $v_t\lesssim 45$ GeV and
		$-0.3\lesssim\sin\alpha\lesssim0.5$ are allowed by the present data.
		If data continue to push the Higgs couplings towards the SM values,
		certain scenarios might still accommodate additional light scalars
		as allowed by the current direct search limits.
		
		\item We have considered the inclusive Higgs cross sections to
		constrain the anomalous Higgs couplings. However, in some cases,
		differential distribution may provide additional information.
		Notably, the degeneracy between $c_t$ and $c_{gg}$ may be lifted
		using differential distribution data
		\cite{Azatov:2013xha,Grojean:2013nya,Banfi:2013yoa}. Admixture of
		CP-odd component to the Higgs can in principle be probed using
		precise analysis of the final state angular distribution
		\cite{Ferreira:2016jea,Bernlochner:2018opw}. On the other hand,
		measurement of the off-shell Higgs production can shed some light on
		the energy dependence of the couplings
		\cite{Azatov:2014jga,Goncalves:2018pkt}.
		
		\item Once the HL-LHC data become available, a better handle on the
		Yukawa couplings, including those involving other fermions
		(e.g. $\tau$ lepton), would unravel even inner layers of underlying
		dynamics. Moreover the future Higgs factories are expected to reduce
		the uncertainties in the Higgs couplings from percentage to
		per-mille level, thus probing deeper into the BSM parameter space.

	\end{itemize}
	
	%%%%%%%%%%%%%%%%%%%%%%%%%%%%%%%%%%%%%%%%%%%%%%%%%%%%%%%%%%%%%%%%%%%%%%%%%%%%%%%%%%%%%%%%%%%%%%%%%%%%%%%%%%%%
	%\newpage
	%\begin{small}
	\section*{Acknowledgments}
	
	A.B. acknowledges support from the Department of Atomic Energy,
	Government of India. G.B. acknowledges support of the J.C. Bose
	National Fellowship from the Department of Science and Technology,
	Government of India (SERB Grant No. SB/S2/JCB-062/2016).
	
	%\end{small}
	
	%%%%%%%%%%%%%%%%%%%%%%%%%%%%%%%%%%%%%%%%%%%%%%%%%%%%%%%%%%%%%%%%%%%%%%%%%%%%%%%%%%%%%%%%%%%%%%%%%%%%%%%%%%%%
	
	\appendix
	
	\section{Higgs signal strength data}
	\label{app_data}
	
	We collect the Higgs signal strength data that we have used to perform
	the $\chi^2$-fitting from the references \cite{Aad:2019mbh,
		CMS:2020gsy, Khachatryan:2016vau, Cepeda:2019klc}, which we tabulate
	below. The covariance matrices are taken from the published papers for
	the ATLAS Run 2 data (Fig.\,6 of \cite{Aad:2019mbh}) and ATLAS + CMS
	combined Run 1 data (Fig.\,27 of \cite{Khachatryan:2016vau}), while for the CMS Run 2 data it is collected from
	the supplementary materials available
	online\footnote{\href{http://cms-results.web.cern.ch/cms-results/public-results/preliminary-results/HIG-19-005/index.html}
		{http://cms-results.web.cern.ch/cms-results/public-results/preliminary-results/HIG-19-005/index.html}}.
	%For the CMS Run 2 data, the covariance matrix is collected from the supplementary materials available online\footnote{\href{http://cms-results.web.cern.ch/cms-results/public-results/preliminary-results/HIG-19-005/index.html}{http://cms-results.web.cern.ch/cms-results/public-results/preliminary-results/HIG-19-005/index.html}}, while the covariance matrices are taken from the published papers for the ATLAS Run 2 data (Fig.\,6 of \cite{Aad:2019mbh}) and ATLAS + CMS combined Run 1 data (Fig.\,27 of \cite{Khachatryan:2016vau}).  
	Further note that for the HL-LHC projections the central values are taken as
	$1.0$ for each of the signal strengths and the uncertainties
	correspond to the projected sensitivities at the end of the HL-LHC
	program (`S2' dataset of \cite{Cepeda:2019klc}). Since the total
	uncertainties for the CMS projections for HL-LHC have not been
	reported, we have added the statistical and systematic parts in
	quadrature.  The uncertainty projections for the future Higgs
	factories are extracted from Table\,5.5 of \cite{Asner:2013psa} for
	ILC, Table\,11 of \cite{An:2018dwb} for CEPC and Table\,4.1 of
	\cite{Abada:2019lih} for FCC-ee.
	
	%\begin{footnotesize}
	\begin{table}[]
		%\begin{footnotesize}
		\centering
		\begin{tabular}{cccccccc}
			\hline\hline \rule{0pt}{2.5ex}	Production & Decay &
			Run 1 & \multicolumn{2}{c}{Run 2} & 
			\multicolumn{2}{c}{HL-LHC \cite{Cepeda:2019klc}}
			\\ \rule{0pt}{2.5ex}	   &	 & ATLAS+CMS \cite{Khachatryan:2016vau} & ATLAS
			\cite{Aad:2019mbh}	& CMS \cite{CMS:2020gsy} &
			ATLAS & CMS\\
			%\rule{0pt}{2.5ex}	 &  & & & & \multicolumn{2}{c}{Central value $\mu=1$} \\
			
			\hline 				
			
			\rule{0pt}{2.5ex}	\multirow{5}{*}{ggH} & $\gamma\gamma$ &
			$1.10\substack{+0.23 \\ -0.22}$
			& $0.96\substack{\pm0.14}$ & $1.09\substack{+0.15 \\ -0.14}$  & $\substack{\pm 0.04}$ &
			$\substack{\pm 0.04}$ \\ 
			
			\rule{0pt}{2.5ex}	& ZZ  & $1.13\substack{+0.34 \\ -0.31}$ & 
			$1.04\substack{+0.16 \\ -0.15}$ & $0.98\substack{+0.12
				\\ -0.11}$ &
			$\substack{\pm 0.04}$ & $\substack{\pm
				0.04}$\\  
			
			\rule{0pt}{2.5ex}	& WW &
			$0.84\substack{\pm 0.17 }$ & $1.08\substack{\pm
				0.19}$ & $1.28\substack{+0.20 \\ -0.19}$ & $\substack{+0.05 \\ -0.04}$ &
			$\substack{\pm 0.03}$\\ 
			
			\rule{0pt}{2.5ex}	&
			$\tau^+\tau^-$ & $1.00\substack{\pm 0.60}$ & $0.96\substack{+0.59 \\ -0.52}$ &
			$0.39\substack{+0.38 \\ -0.39}$  &
			$\substack{+0.12 \\ -0.11}$ & $\substack{\pm 0.06}$
			\\ 
			
			\rule{0pt}{2.5ex}	& $b\bar{b}$ & -- & -- &
			$2.45\substack{+2.53 \\ -2.35}$  & -- & --\\	 
			
			\hline
			
			\rule{0pt}{2.5ex}	\multirow{5}{*}{VBF}  & $\gamma\gamma$ & $1.30\substack{\pm 0.50}$
			& $1.39\substack{+0.40 \\ -0.35}$ & $0.77\substack{+0.37
				\\ -0.29}$  & $\substack{+0.10
				\\ -0.09}$ & $\substack{\pm 0.14}$\\ 
			
			\rule{0pt}{2.5ex} & ZZ & $0.10\substack{+1.10 \\ -0.60}$
			& $2.68\substack{+0.98 \\ -0.83}$ & $0.57\substack{+0.46
				\\ -0.36}$  &
			$\substack{\pm 0.12}$ & $\substack{\pm
				0.18}$\\  
			
			\rule{0pt}{2.5ex}	& WW &
			$1.20\substack{\pm 0.40}$ & $0.59\substack{+0.36
				\\ -0.35}$ & $0.63\substack{+0.65 \\ -0.61}$  & $\substack{+0.10 \\ -0.09}$ &
			$\substack{\pm 0.09}$\\ 
			
			\rule{0pt}{2.5ex}		&
			$\tau^+\tau^-$ & $1.30\substack{\pm 0.40}$ & $1.16\substack{+0.58 \\ -0.53}$ &
			$1.05\substack{+0.30 \\ -0.29}$  &
			$\substack{\pm0.08}$ & $\substack{\pm
				0.06}$\\ 
			
			\rule{0pt}{2.5ex}		& $b\bar{b}$ & -- &
			$3.01\substack{+1.67 \\ -1.61}$ & -- & -- & --\\	 
			
			\hline
			
			\rule{0pt}{2.5ex}	\multirow{5}{*}{VH} & $\gamma\gamma$ & -- &
			$1.09\substack{+0.58 \\ -0.54}$ & -- & $\substack{\pm
				0.09}$ & --\\ 
			
			\rule{0pt}{2.5ex}		& ZZ & -- &
			$0.68\substack{+1.20 \\ -0.78}$ & -- & $\substack{+0.19
				\\ -0.18}$ & --\\  
			
			\rule{0pt}{2.5ex}	    & WW & -- & -- &
			-- & -- & --\\ 
			
			\rule{0pt}{2.5ex}		&
			$\tau^+\tau^-$ & -- & -- & -- & -- & --\\ 
			
			\rule{0pt}{2.5ex} &
			$b\bar{b}$ & -- & $1.19\substack{+0.27 \\ -0.25}$ & -- & -- &
			$\substack{\pm 0.06}$\\
			
			\hline
			
			\rule{0pt}{2.5ex}	\multirow{5}{*}{WH} 	&
			$\gamma\gamma$ & $0.50\substack{+1.30 \\ -1.20}$ & -- & -- &
			-- & $\substack{\pm 0.20}$\\ 
			
			\rule{0pt}{2.5ex}		& ZZ & -- &
			-- & $1.10\substack{+0.96 \\ -0.74}$ & -- &
			$\substack{\pm 0.67}$\\  
			
			\rule{0pt}{2.5ex}	    & WW & $1.60\substack{+1.20
				\\ -1.00}$ & --
			& $2.85\substack{+2.11 \\ -1.87}$  & -- & $\substack{\pm 0.19}$\\ 
			
			\rule{0pt}{2.5ex}
			& $\tau^+\tau^-$ &
			$-1.4\substack{\pm 1.4}$ & -- & $3.01\substack{+1.65 \\ -1.51}$ & -- & --\\ 
			
			\rule{0pt}{2.5ex} &
			$b\bar{b}$ &
			$1.00\substack{\pm 0.50}$ & -- & $1.27\substack{+0.42 \\ -0.40}$ & $\substack{\pm0.10}$ & --\\	 
			
			\hline
			
			\rule{0pt}{2.5ex}	\multirow{5}{*}{ZH} 	&
			$\gamma\gamma$ & $0.50\substack{+3.00 \\ -2.50}$  & -- & -- &
			-- & $\substack{\pm 0.33}$\\ 
			
			\rule{0pt}{2.5ex}		& ZZ & -- &
			-- & $1.10\substack{+0.96 \\ -0.74}$ & -- &
			$\substack{\pm 1.09}$\\  
			
			\rule{0pt}{2.5ex}	    & WW & $5.90\substack{+2.60
				\\ -2.20}$ & --
			& $0.90\substack{+1.77 \\ -1.43}$ & -- & $\substack{\pm 0.25}$\\ 
			
			\rule{0pt}{2.5ex}
			& $\tau^+\tau^-$ &
			$2.20\substack{+2.20 \\ -1.80}$ & -- & $1.53\substack{+1.60 \\ -1.37}$ & -- & --\\ 
			
			\rule{0pt}{2.5ex}
			& $b\bar{b}$ &
			$0.40\substack{\pm 0.40}$ & -- & $0.93\substack{+0.33 \\ -0.31}$ & $\substack{\pm0.45}$ & --\\	 
			
			\hline 
			
			\rule{0pt}{2.5ex}
			\multirow{5}{*}{$t\bar{t}H+tH$} &
			$\gamma\gamma$ & -- & $1.10\substack{+0.41
				\\ -0.35}$ & -- & -- &
			--\\  
			
			\rule{0pt}{2.5ex} & VV & -- &
			$1.50\substack{+0.59 \\ -0.57}$ & -- & --
			& --\\ 
			
			\rule{0pt}{2.5ex}		&
			$\tau^+\tau^-$ & -- & $1.38\substack{+1.13
				\\ -0.96}$ & -- & -- &
			--\\ 
			
			\rule{0pt}{2.5ex}		& $b\bar{b}$ & -- &
			$0.79\substack{+0.60 \\ -0.59}$ & -- & --
			& --\\
			
			\hline 
			
			\rule{0pt}{2.5ex}	\multirow{5}{*}{$t\bar{t}H$}  &
			$\gamma\gamma$ &
			$2.20\substack{+1.60 \\ -1.30}$ & -- & $1.62\substack{+0.52 \\ -0.43}$ & $\substack{+0.08 \\ -0.07}$
			& $\substack{\pm 0.11}$\\ 
			
			\rule{0pt}{2.5ex}	& ZZ & -- & -- &
			$0.25\substack{+1.03 \\ -0.25}$ & $\substack{+0.23
				\\ -0.20}$ &  $\substack{\pm 0.33}$\\  
			
			\rule{0pt}{2.5ex} &
			WW &
			$5.00\substack{+1.80 \\ -1.70}$ & -- & $0.93\substack{+0.48 \\ -0.45}$  & -- & -- \\ 
			
			\rule{0pt}{2.5ex}
			& $\tau^+\tau^-$ &
			$-1.9\substack{+3.7 \\ -3.3}$ & -- & $0.81\substack{+0.74 \\ -0.67}$ & -- & -- \\ 
			
			\rule{0pt}{2.5ex} &
			$b\bar{b}$ &
			$1.10\substack{\pm 1.00}$ & -- & $1.13\substack{+0.33 \\ -0.30}$ & -- & -- \\	 
			
			\hline\hline 
			
		\end{tabular} 
		\caption{\small\it The Higgs signal strength data used to
			perform the $\chi^2$-analysis. For the HL-LHC case, only the
			projected uncertainties are reported assuming the central
			values to be equal to $1.0$.}
		\label{data}
		%\end{footnotesize}	
	\end{table}
	%\end{footnotesize}

%%%%%%%%%%%%%%%%%%%%%%%%%%%%%%%%%%%%%%%%%%%%%%%%%%%%%%%%%%%%%%%%%%%%%%%%%%%%%%%%%%%%%%%%%%%%%%%%

\newpage
\bibliographystyle{JHEP} \bibliography{Higgs_globalfit}

%%%%%%%%%%%%%%%%%%%%%%%%%%%%%%%%%%%%%%%%%%%%%%%%%%%%%%%%%%%%%%%%%%%%%%%%%%%%%%%%%%%%%%%%%%%%%%%%%%
\end{document}